\documentclass[useAMS,usenatbib]{mn2e}
\usepackage[dvips]{graphicx}
\usepackage{hyperref}
\usepackage{epstopdf}
\usepackage{times}
\usepackage{amsmath}
\usepackage{amssymb}
\usepackage{color}
\newcommand{\aap}{A\&A}

\newcommand{\mnras}{MNRAS}

\newcommand{\apj}{ApJ}
\newcommand{\apjl}{ApJ}

\newcommand{\pasj}{PASJ}

\newcommand{\aj}{AJ}

\newcommand{\prd}{Phys. Rev. D}

\hypersetup{colorlinks,citecolor=blue,linkcolor=blue}

\newcommand{\msun}{${\rm M}_{\odot}$}
\title[BBH mergers from GCs with IMBHs]{{\sc mocca-survey} Database I: Binary Black Hole Mergers from Globular Clusters with Intermediate Mass Black Holes}
\author[J. Hong et al.]  {Jongsuk Hong,$^{1,2}$\thanks{hongjs@yonsei.ac.kr (JH); sjyoon0691@yonsei.ac.kr (SJY)} 
Abbas Askar,$^{3}$ Mirek Giersz,$^4$ Arkadiusz Hypki$^5$ \& Suk-Jin Yoon$^{1\star}$
\\
  $^1$ Department of Astronomy, Yonsei University 50 Yonsei-Ro, Seodaemun-Gu, Seoul 03722, Republic of Korea\\
  $^2$ Kavli Institute for Astronomy and Astrophysics, Peking University, Yi He Yuan Lu 5, HaiDian District, Beijing 100871, China\\
  $^3$ Lund Observatory, Department of Astronomy, and Theoretical Physics, Lund University, Box 43, SE-221 00 Lund, Sweden\\
  $^4$ Nicolaus Copernicus Astronomical Centre, Polish Academy of Sciences, ul. Bartycka 18, 00-716 Warsaw, Poland\\
  $^5$ Astronomical Observatory Institute, Faculty of Physics, A. Mickiewicz University, S\l{}oneczna 36, 60-286 Pozna{\'n}, Poland
}
\begin{document}

\date{Accepted 2020 August 21. Received 2020 August 20; in original form 2020 June 1}
\maketitle

\label{firstpage}

\begin{abstract}
The dynamical formation of black hole binaries in globular clusters that merge due to gravitational waves occurs more frequently in higher stellar density. Meanwhile, the probability to form intermediate mass black holes (IMBHs) also increases with the density. To explore the impact of the formation and growth of IMBHs on the population of stellar mass black hole binaries from globular clusters, we analyze the existing large survey of Monte-Carlo globular cluster simulation data ({\sc mocca-survey} Database I). We show that the number of binary black hole mergers agrees with the prediction based on clusters' initial properties when the IMBH mass is not massive enough or the IMBH seed forms at a later time. However, binary black hole formation and subsequent merger events are significantly reduced 
compared to the prediction when the present-day IMBH mass is more massive than $\sim10^4 \rm M_{\odot}$ or the present-day IMBH mass exceeds about 1 per cent of cluster's initial total mass. By examining the maximum black hole mass in the system at the moment of black hole binary escaping, we find that $\sim$ 90 per cent of the merging binary black holes escape before the formation and growth of the IMBH. Furthermore, large fraction of stellar mass black holes are merged into the IMBH or escape as single black holes from globular clusters in cases of massive IMBHs, which can lead to the significant under-population of binary black holes merging with gravitational waves by a factor of 2 depending on the clusters' initial distributions. 
\end{abstract}


\section{Introduction\label{S1}}
The first direct gravitational wave detection has been made in 2015 from a merger of two black holes in a binary system \citep{2016PhRvL.116f1102A} after Einstein's prediction a century ago. So far, a dozen GW detections were announced in the first two observing runs of the LIGO-VIRGO detector network \citep{2019PhRvX...9c1040A}, which are all in forms of coalescence of compact binaries composed of black holes or neutron stars \citep[mostly black hole binaries but a couple of neutron star binaries including GW170817 and GW190425\footnote{The probability that it is a black hole and neutron star binary is still not ruled out};][]{2017PhRvL.119p1101A,2020ApJ...892L...3A}. There have been several mechanisms suggested for the formation of black hole binaries as gravitational wave sources: binary stellar evolution of isolated binaries \citep{2002ApJ...572..407B,2007ApJ...662..504B,2012ApJ...759...52D}, three-body binary formation in the dense stellar systems \citep{2000ApJ...528L..17P,2006ApJ...637..937O,2010MNRAS.402..371B,2011MNRAS.416..133D,2014MNRAS.440.2714B,2015ApJ...800....9M,2017MNRAS.469.4665P,2019MNRAS.486.3942K,2019Natur.576..406S,2019MNRAS.487.2947D,2020arXiv200610771K}, gravitational radiation capture \citep{2009MNRAS.395.2127O,2015MNRAS.448..754H,2017PhRvD..96h4009B,2018MNRAS.481.5445S,2020PhRvD.101l3010S} and Kozai-Lidov mechanism in hierarchical triple systems \citep{2012MNRAS.422..841A,2012ApJ...757...27A,2018ApJ...860....5G,2018ApJ...863....7R}.
In the context of formation environments, the field formation is considered as the most dominant contributor \citep[e.g.][]{2016ApJ...819..108B,2020A&A...636A.104B} but the dynamical formation in dense environments becomes more important as gravitational wave events with possibilities of repeated mergers were discovered \citep{2020RNAAS...4....2K,2020ApJ...896L..10R}. 
Many more detections are expected during the next observing run due to huge improvements in detection sensitivity and cooperation of multiple detectors over the world. 
Since different formation channels can cause different physical properties of coalescing binaries, such as spin, total mass, mass ratio and orbital eccentricities \citep{2015MNRAS.448..754H,2016MNRAS.458.3075A,2018MNRAS.481.5445S,2018PhRvD..98l3005R,2019ApJ...886...25B,2020ApJ...894..133A,2020A&A...636A.104B}, accumulation of detection will be able to shed light on the formation history of gravitational wave sources. 

Dense stellar systems such as globular clusters and nuclear star clusters in galactic centres are very efficient birth places for gravitational wave sources in spite of their relatively small mass fraction in the cosmological star formation history. Dynamical interactions in the high density regions can form binary black holes and also can drive them to be tightened until these binaries enter the relativistic regime \citep{2019PhRvD.100d3010S}. A number of numerical studies have confirmed the role of internal dynamics in the formation and evolution of binary black hole as gravitational wave sources \citep{2015ApJ...800....9M,2017MNRAS.464L..36A,2018MNRAS.480.5645H}. Especially, \citet{2018MNRAS.480.5645H} found an empirical relation between merging binary black hole production and the host globular clusters' properties, based on a large set of Monte-Carlo cluster simulations. They suggested that the number of binary black hole mergers is closely correlated to the cluster's mass, half-mass radius and binary fraction. These findings can give us a clue for the link between the gravitational wave detection in this epoch and the formation of globular clusters in the early Universe \citep[see also ][]{2017PASJ...69...94F,2019ApJ...873..100C,2019MNRAS.487....2M}.

However, it is also possible for a very dense stellar system to form an intermediate-mass black hole (IMBH). Since early 2000s, there have been lots of theoretical and observational efforts to prove the presence of IMBHs in globular clusters \citep[e.g.][]{2002MNRAS.330..232C,2002ApJ...576..899P,2005ApJ...620..238B,2005ApJ...634.1093G,2010ApJ...710.1063V}. The most probable formation channel suggested by theoretical studies is the runaway collisions of massive stars in early globular clusters. The collisions can lead to a rapid formation of very massive object, which can grow to an IMBH by mergers with other stars \citep{2004Natur.428..724P,2017MNRAS.472.1677S,2019MNRAS.484.4665S}. More recently, \citet{2015MNRAS.454.3150G} numerically confirmed two IMBH formation scenarios. One is the {\it fast} scenario which corresponds to the runaway collision scenario as mentioned above. The other one is the {\it slow} scenario \citep[for the review, see][]{2019arXiv191109678G}. In this scenario, IMBHs form later by mergers of black holes in the post core-collapse phase of long-term evolution of globular clusters \citep[see also][]{2019MNRAS.486.5008A}. This can happen in less dense clusters than those where the IMBH formation occurs via runaway collisions. 
Radio observations looking for accretion signatures of an IMBH in Galactic globular clusters have found no evidence for their presence \citep{2018ApJ...862...16T}. Previously, the discussion on the presence of IMBHs relied on indirect evidence from the kinematic structure of stars at the central regions of globular clusters \citep[e.g.][]{2002AJ....124.3270G,2010ApJ...710.1032A,2011A&A...533A..36L,2013ApJ...769..107L,2017MNRAS.464.3090A,2017MNRAS.467.4057D}. However, a recent tidal disruption event is expected to open a new window to search and prove IMBHs in globular clusters \citep{2018NatAs...2..656L} in the interim before the construction and operation of deci-Hz gravitational wave detectors which will target intermediate mass ratio inspirals \citep[see][ and references therein]{2019arXiv190811375A}.

In this paper, we explore the effects of the formation and growth of IMBHs on the dynamical formation of stellar mass black hole binaries from globular clusters as progenitors of gravitational wave events. The existence of a massive black hole at the central region of a stellar system can affect the dynamical evolution of the host system overall. The kinetic energy generated by the gravitational interactions between the massive black hole and stars at the vicinity can make the whole cluster expand \citep{2004ApJ...613.1133B,2004ApJ...613.1143B}. The presence of a massive black hole at the center can prevent the core collapse, and the three-body binary formation is also suppressed. The mass segregation procedure also becomes inefficient with an IMBH as noted by \citet{2004ApJ...613.1143B}. Moreover, strong gravity of an IMBH can disrupt binaries around the sphere of influence \citep{2007MNRAS.374..857T}. Since all these processes are very crucial for the formation and hardening of black hole binaries to become gravitational wave merger progenitors, the presence of the IMBH is expected to play a significant role in shaping the population of black hole binaries as gravitational wave sources.

The paper is organized as follows. In Section 2 we describe the numerical code and globular cluster models used in this work.
In Section 3, we present the effects of IMBHs on the binary black hole merger population. We investigate, in Section 4, the underlying astrophysics that induces the phenomena observed in the simulations and propose a new correlation including the effects of IMBHs. Finally, we draw our conclusions in Section 5.\\

\section{Methods and Models\label{S2}}
In this study, we made use of the results for around two thousand star cluster models that
were simulated with the \textsc{mocca} code \citep{2013MNRAS.429.1221H,2013MNRAS.431.2184G} for star cluster
evolution as part of the \textsc{mocca-survey} Database I \citep{2017MNRAS.464L..36A}. The
\textsc{mocca} code is based on Michel H\'enon's Monte Carlo method
\citep{1971Ap&SS..14..151H,1971Ap&SS..13..284H} which was further improved by \citet{1982AcA....32...63S,1986AcA....36...19S} and
\citet{1998MNRAS.298.1239G,2001MNRAS.324..218G,2013MNRAS.431.2184G}. This method combines a statistical treatment for
relaxation with the particle based approach of direct \textit{N}-body simulations to
follow the long term dynamical evolution of spherically symmetric stellar clusters on a
timescale which is a fraction of the cluster's relaxation time. The particle based
approach of this method allows for inclusion of a number of physical processes that are
important for following the evolution of a realistic globular cluster. The \textsc{mocca}
code incorporates stellar and binary evolution prescriptions from the \textsc{SSE}
\citep{2000MNRAS.315..543H} and \textsc{BSE} \citep{2002MNRAS.329..897H} codes. For computing the outcome of
strong dynamical binary-single and binary-binary encounters, \textsc{mocca} use the
\textsc{fewbody} code \citep{2004MNRAS.352....1F} which is a direct \textit{N}-body code for
simulating small-\textit{N} gravitational dynamics and carrying out scattering
experiments. Additionally, the code also incorporates the escape process in a static tidal
field based on prescriptions from \citet{2000MNRAS.318..753F}. For the tidal field, a point mass
approximated Galactic potential is assumed and the star cluster is placed on a circular
orbit at a given Galactocentric radius.  

The main advantage of the \textsc{mocca} code and the Monte Carlo method is
computational speed, a realistic globular cluster with more than a million stars with a high
initial binary fraction can be simulated up to a Hubble time within a week. Detailed
comparison between results of evolved star cluster models simulated with the
\textsc{mocca} code and direct \textit{N}-body simulations show remarkable agreement for
the evolution of both global properties and populations of specific objects and stars
\citep{2003MNRAS.343..781G,2016MNRAS.458.1450W,2017MNRAS.470.1729M,2019MNRAS.487.2412G}.

Due to the speed of the \textsc{mocca} code, it is well suited for parameter space
exploration and checking how changes in initial conditions for globular cluster models can
influence the formation and retention of specific populations of stars and compact objects.
Additionally, the influence of those initial conditions on the long term dynamical
evolution of the cluster can also be thoroughly investigated with these results. The
\textsc{mocca-survey} Database I \citep{2017MNRAS.464L..36A} project is a collection of about two
thousand evolved star cluster models with different initial parameters that include the number
of stars, fraction of binary systems, distribution of initial binary parameters,
metallicity, central density, half-mass radii, tidal radii, and different assumptions for
natal kicks of black holes.

The initial conditions of the simulated star cluster models are provided in Table 1 in \citet{2017MNRAS.464L..36A}.
These models have initial number of objects (single stars and binary
systems) of $4 \times 10^{4}$, $1 \times 10^{5}$, $4 \times 10^{5}$, $7 \times 10^{5}$ and
$1.2 \times 10^{6}$. The initial binary fraction for the models were 5, 10, 30 and 95 per
cent. All these models were initially \citet{1966AJ.....71...64K} models with initial concentration
parameter (\rm $W_{0}$) values of 3, 6 and 9. For each model, we had used initial tidal
radius values of 30, 60 and 120 pc. Depending on the combination of cluster mass and tidal
radius, the Galactocentric distances for these models ranged between 1 kpc to up to about
50 kpc. The half-mass radii of tidally underfilling models ranged from 0.6 pc to 4.8 pc.
A smaller number of tidally-filling models were also simulated in which half-mass radius
values ranged from several to up to few tens of pc. For models with 5, 10 and 30 per cent
primordial binary fraction, the semi-major axis distribution is uniform in logarithm with
values of up to 100 AU. The initial binary eccentricity distribution is thermal and mass
ratio distribution is uniform. For models with 95 per cent primordial binary fraction, the
initial binary parameters are set according to the distributions provided by
\citet{1995MNRAS.277.1491K,1995MNRAS.277.1507K}. For each cluster model, the zero age main sequence (ZAMS)
mass of stars was sampled between 0.08 $\rm M_{\odot}$ to 100 $\rm M_{\odot}$ according to
the two-component initial mass function (IMF) given by \citet{2001MNRAS.322..231K}. The metallicity
(Z) values for the simulated models were 0.0002, 0.001, 0.005, 0.006 and 0.02, with 0.001
being the most frequently used value. For all models, neutron star natal kick velocity
follows a Maxwellian distribution (with $\sigma = 265\,\text{km }\text{s}^{-1}$) based on
observations of pulsar proper motions \citep{2005MNRAS.360..974H}. For black holes, nearly a half of
the models uses the same natal kick distribution as neutron stars and for the other half,
the natal kicks are modified according to the mass fallback prescription provided by
\citet{2002ApJ...572..407B}. This prescription leads to higher average of mass of black holes
and also reduces their natal kicks which results in higher retention fractions in the
simulated cluster models \citep{2018MNRAS.479.4652A,2018MNRAS.478.1844A}. \\

\section{Results}
\subsection{{\sc mocca-survey} cluster models}
\begin{figure}
  \centering
  \includegraphics[width=1.0\columnwidth]{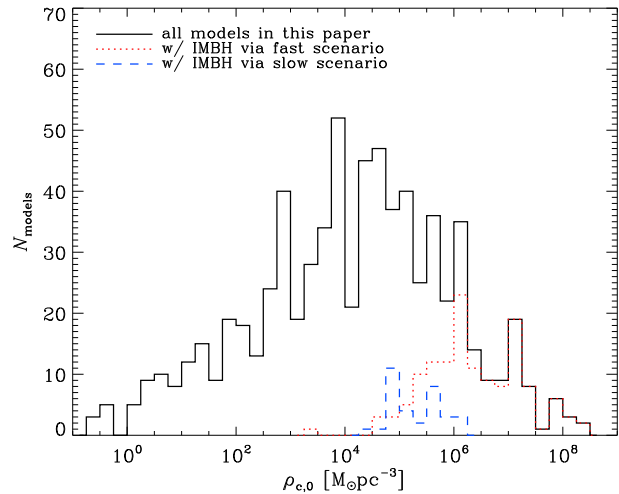}
  \caption{Distribution of central density for the models considered in this study. Red and blue histograms represent the distribution of models with an IMBH via the fast and slow formation scenarios, respectively.}\label{F1}
\end{figure}
\begin{figure}
  \centering
  \includegraphics[width=1.0\columnwidth]{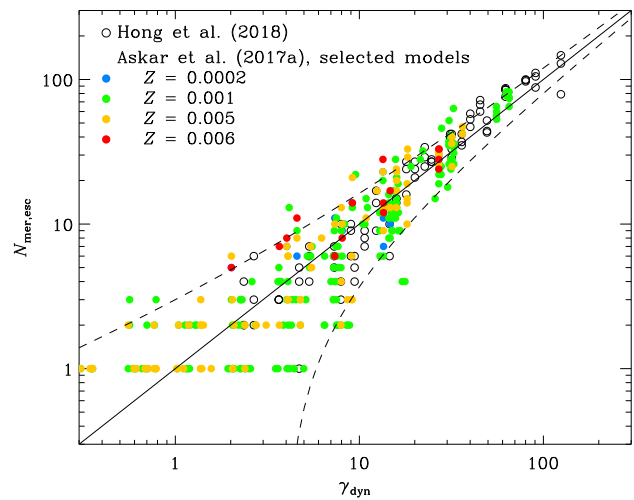}
  \caption{The actual number of escaping merging binary black holes as a function of $\gamma_{\rm dyn}$ parameter which gives the expected number of merging black holes formed dynamically in globular clusters. The definition of $\gamma_{\rm dyn}$ is taken from \citet{2018MNRAS.480.5645H}. Open circles correspond to the simulation models used in \citet{2018MNRAS.480.5645H} and filled circles are for the {\sc mocca-survey} models selected in this paper from \citet{2017MNRAS.464L..36A}. Note, however, that {\sc mocca-survey} models with an IMBH are excluded in this figure for direct comparison. Dashed lines represent 2 times of the Poisson error in numbers. Different colors represent the metallicity.}\label{F2}
\end{figure}
\begin{figure}
  \centering
  \includegraphics[width=1.0\columnwidth]{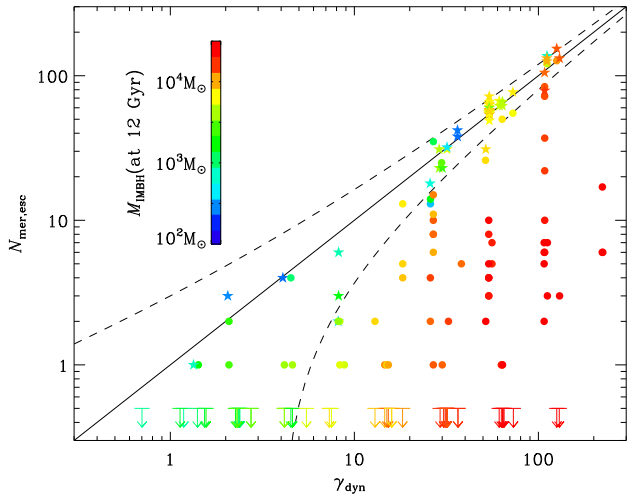}
  \caption{Same as Fig. 2 but only for the modes with the IMBH formation. Different colors indicate the IMBH mass at 12 Gyr and different symbols show the formation scenario either fast (circle) or slow (star). Downward arrows represent models with an IMBH but no escaping merger.}\label{F3}
\end{figure}
In order to investigate the effects of intermediate mass black hole formation on the binary black hole merger events, we first reduce the survey models having similar conditions and assumptions. 
For example, \citet{2017MNRAS.464L..36A} found that the number of mergers strongly depends on the fallback during the supernova explosion which determines the final mass of the black hole and influences the natal kick that the black hole receives. Models with fallback prescriptions lead to lower kicks and higher probability of black hole retention in the cluster. In this paper, we limit our focus to the models with the fallback prescription considered. The metallicity is also important by affecting the wind mass-loss at the giant phases. \citet{2017MNRAS.464L..36A} also revealed that there is a discrepancy in the population of binary black hole mergers between solar and sub-solar metallicity \citep[for more details on the metallicity dependence of the merger population, see e.g.][]{2018MNRAS.474.2959G,2017ApJ...836L..26C,2020MNRAS.tmp.2334D}. Thus, we finally select 702 models with Z $=$ 0.0002, 0.001, 0.005 and 0.006 lower than the solar metallicity (Z$_{\odot} =$ 0.02). But we take full variety of other parameters such as the number of stars, tidal radius, tidal filling factor, dimensionless \citet{1966AJ.....71...64K} concentration parameter $W_0$, and the binary fraction as mentioned in Section 2. 

We start our presentation by showing the distribution of models used in the analysis. 
Fig. \ref{F1} shows the number of models used for the investigation in this paper as a function of initial central density.
The number of models peaks at the middle range of the initial central density ($\sim10^4$ ${\rm M}_{\odot}{\rm pc}^{-3}$). In this figure, we also show the central density distribution of models hosting an IMBH which have grown by the fast or slow scenarios. We assume that all globular clusters form 12 Gyr ago, and define models with IMBHs by whether the most massive object at 12 Gyr exceeds 200 \msun. The formation scenario is determined by the maximum black hole mass at 1 Gyr \citep[see also section 2.1 in][]{2019arXiv191109678G}. If the IMBH seed black hole mass becomes more massive than 200 \msun\footnote{This criterion for the IMBH formation is an arbitrary value currently. According to \citet{2017MNRAS.470.4739S}, the black hole mass from the stellar evolution of very massive stars at low metallicity can be larger than 200 \msun. We plan to develop a better classification between the fast and slow formation cases by upgrading stellar/binary evolution recipes.} before 1 Gyr, we classify these models as fast formation cases. On the other hand, the models whose seed IMBH black holes form after 1 Gyr are regarded as slow formation cases. Under this definition, there are 117 fast formation cases and 43 slow formation cases. It is clear from Fig. \ref{F1} that the IMBH formation by fast scenario needs higher central density than that by the slow scenario. 
\begin{figure}
  \centering
  \includegraphics[width=1.0\columnwidth]{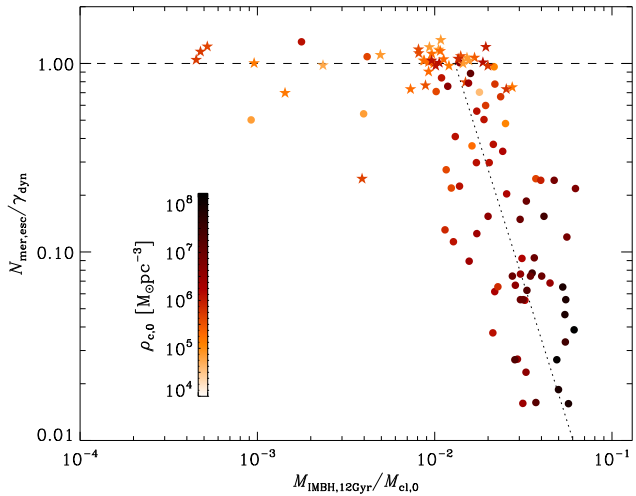}
  \caption{The ratio of escaping black hole binary mergers to the expectation as a function of the IMBH mass fraction with respect to the initial cluster mass. Symbols are the same as Fig. \ref{F3} but colors represent the initial central density. The dotted line is not a fitting but to show the trend.}\label{F4}
\end{figure}

\subsection{IMBH and stellar mass black hole merger population}
Before investigating the effects of IMBHs, we need to quantify the expected number of binary black hole mergers for all survey models. 
There is a sophisticated correlation formula between the cluster parameters and the number of mergers among the escaping binary black holes \citep{2018MNRAS.480.5645H}. 
They found that the number of black hole binaries which form and escape dynamically through binary encounters and eventually merge within 12 Gyr is not sensitive to the tidal field, binary fraction and distribution of host globular clusters, but tightly correlated with the initial total mass and the half-mass density by following
\begin{equation}\label{E1}
\gamma_{\rm dyn}\equiv A\cdot \frac{M_{0}}{10^{5}{\rm M}_{\odot}}\times \Big(\frac{\rho_{\rm h}}{10^{5}{\rm M}_{\odot}{\rm pc}^{-3}}\Big)^{\alpha},
\end{equation}
with fitting coefficients $A=12.3$ and $\alpha=0.33$\footnote{Note that in their study, in-cluster mergers are not included in the derivation of the correlation. The correlation formula can be different when including in-cluster mergers \citep[see e.g.][]{2020MNRAS.492.2936A}.} where $M_{0}$ and $\rho_{\rm h}$ are the initial total mass and the half-mass density.
Fig. \ref{F2} manifests how this correlation works well for the {\sc mocca-survey} models without the IMBH formation. Interestingly, {\sc mocca-survey} models have three choices of \citet{1966AJ.....71...64K} concentration $W_0$, but, as shown in the figure, there is no significant spread of correlations against the concentration compared to \citet{2018MNRAS.480.5645H} which used one concentration $W_0=7$. We also note from Fig. \ref{F2} that there is no major effects of the different metallicity on the correlations. Then, we apply this correlation to the {\sc mocca-survey} models that form IMBHs during their lifetime.

In Fig. \ref{F3}, we show that the number of escaping merging binary black holes in the models with an IMBH is smaller than that without the IMBH. The degree of discrepancy becomes larger for the models with more massive IMBHs. For instance, when the IMBH mass at 12 Gyr is greater than $\sim2\times10^4$ \msun, the number of mergers is reduced by a factor of 10 overall compared to the prediction. Another interesting remark is that for the cases with slow IMBHs, there is no significant difference in the expected number of mergers with respect to non-IMBH cases. 

Fig. \ref{F4} interprets the impact of the IMBH formation and its mass buildup on the binary black hole merger production. This figure shows the ratio of the number of mergers to the prediction from the formula as a function of the final mass fraction (at 12 Gyr) of IMBHs compared to the clusters' initial mass. While the number of mergers is similar for the cases with the final IMBH mass less than 1 per cent of the initial cluster mass, the ratio systematically drops as the final IMBH mass fraction becomes larger. We note that the the final IMBH mass fraction and the reduction of merger population become more significant as the initial central density increases. It is also consistent with the finding in Fig. \ref{F3} that models with the fast IMBH formation produce less mergers than non- or slow IMBH formation cases. Therefore, there might be some discrepancies in the dynamical history for IMBH growth between fast and slow cases. In contrast with fast IMBH formation cases, the seed IMBHs for slow formation cases usually form around the core collapse of the entire system when nearly all black holes have already escaped \citep{2015MNRAS.454.3150G}. Thus, the escaping binary black holes and subsequent mergers are not influenced by the presence of IMBHs in their formation, which will be discussed in the next section.

\section{Discussion}
\subsection{How do IMBHs affect the merger population?}
\begin{figure}
  \centering
  \includegraphics[width=1.0\columnwidth]{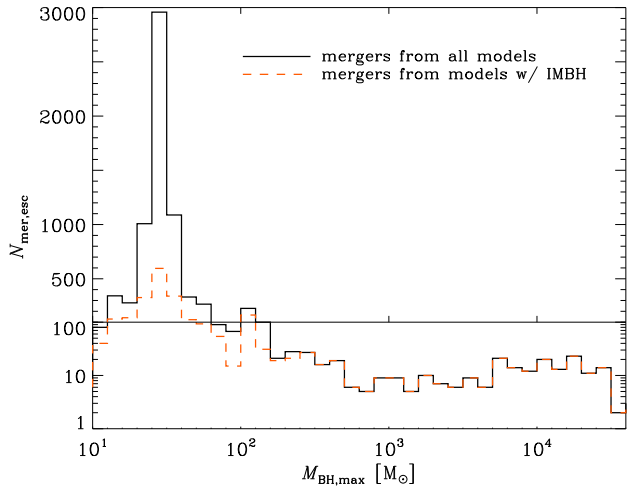}
  \caption{Number distribution of the most massive object in the cluster at the escaping time of merging black hole binaries formed dynamically. The peak locates around 30 \msun. About 10 per cent of binary black holes escape when the maximum mass object is more massive than 200 \msun. Dashed line presents the same distribution but from the cluster models which host IMBHs at 12 Gyr.}\label{F5}
\end{figure}

Since most of escaping binary black holes were produced and ejected from stellar systems at the very beginning of cluster evolution \citep{2017MNRAS.464L..36A,2018MNRAS.480.5645H}, the presence of an IMBH at 12 Gyr does not necessarily mean that binary black holes were affected by the IMBH at the moment of escape. 
In order to understand the discrepancy in the binary black hole merger production between the fast and slow scenarios, we present the mass of the most massive object in the system when black hole binaries escape from the system in Fig. \ref{F5}. As shown in this figure, almost all (2171 out of 2377) black hole binaries escape when the most massive object is less massive than 200 M$_{\odot}$, but still there are few black hole binaries ejected from the system with very massive IMBHs ($>10^3$ \msun). It is obvious that binary black holes can escape under the presence of IMBHs, but even in these cases, most of binary black holes escape before the formation and growth of the IMBH.

\begin{figure}
  \centering
  \includegraphics[width=1.0\columnwidth]{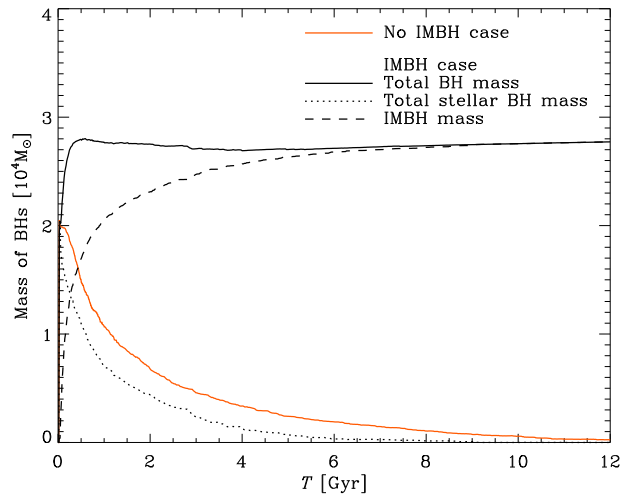}
  \caption{Sum of black hole masses for two representative models with (black lines) and without (orange line) the IMBH formation. The black solid line represents the total mass of all black holes in the clusters. The black dotted and dashed lines show the mass of stellar mass black holes and of the IMBH, respectively.}\label{F6}
\end{figure}

In \citet{2018MNRAS.480.5645H}, some possible explanations were proposed for the effects of IMBHs on the black hole binary formation: depletion of massive main sequence stars by runaway mergers, prevention of binary black hole formation by the strong gravity, and ejection of black holes as singles after gravitational interactions at the vicinity of IMBHs. To test this idea, we compare two models whose simulation parameters are almost identical but one forms an IMBH while the other does not. We select two clusters with the same simulation parameters except for the different concentration $W_0=$ 3 and 9. According to prediction based on the equation \ref{E1}, they are supposed to produce similar numbers of escaping merging binary black holes, which is 118. However, the actual numbers of mergers are 7 for the IMBH case ($W_0=$ 9) and 121 for the non-IMBH case ($W_0=$ 3). 

Fig. \ref{F6} shows the total mass of black holes, either an IMBH or stellar mass black holes. The total mass of black holes directly after the stellar evolution of progenitor stars has finished is similar between two cases. The number of black holes formed via stellar evolution is not very different between two cases, which infers that the first hypothesis, the depletion of massive main sequence stars by runaway mergers, from \citet{2018MNRAS.480.5645H} is not the case. For the non-IMBH case, the total black hole mass remaining in the system quickly reaches a peak and decreases continuously as black holes escape from the system. On the other hand, for the model with an IMBH, the IMBH mass immediately goes to over $10^4$ \msun{ }in 50 Myr as a consequence of runaway mergers into the seed IMBH. The total mass of stellar mass black holes rapidly decreases until $\sim$5 Gyr, but the total black hole mass remains nearly constant. It indicates that most of stellar mass black holes have been eaten up by the IMBH instead of escaping. In the same time, the IMBH mass keeps growing and well exceeds the initial black hole mass via stellar evolution ($\sim2\times10^4$ \msun).   After 5 Gyr, the IMBH growth slows down, and only tiny fraction of stellar mass black holes remains in the cluster.
\begin{figure}
  \centering
  \includegraphics[trim=15 10 5 5,width=1.0\columnwidth]{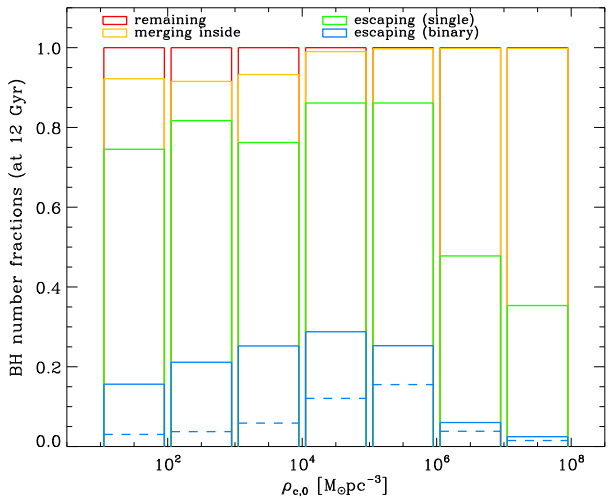}
  \caption{The number fraction of black holes in different populations at 12 Gyr as a function of the central density. We accumulate all black holes from corresponding models.  Red, orange, green and blue boxes represent the fraction of black holes in forms of remainders, in-cluster mergers, single escapers and binary escapers, respectively. Dashed lines show the fractions of binary black hole escapers merging via gravitational waves within a Hubble time.}\label{F7}
\end{figure}

We describe the fraction of black hole populations at 12 Gyr as a function of the central density at the beginning in Fig. \ref{F7}. The number of escaping black holes is significantly reduced when there is a massive IMBH in a cluster.
We note that, in the context of binary black hole mergers as gravitational wave sources, the binary fraction among the escaping black holes is also important as much as the absolute number of escaping black holes. We sum up entire black holes from all the models in the corresponding central density ranges. For the low density cases, approximately 10 per cent of black holes still remain in the cluster, which might be the leftovers of black hole subsystems \citep{2018MNRAS.479.4652A,2018MNRAS.478.1844A,2019MNRAS.487.2412G}. About 20 per cent of the black holes disappear as they merge with another black holes by the binary stellar evolution or gravitational wave events. 3/4 of black holes escape from the system, 20 per cent of the escaping binaries are in binaries and 20 per cent of the escaping black hole binaries eventually merge subsequently with gravitational waves. The fraction of remaining black holes decreases with increasing central density. Both the escaping fraction and the binary fraction among the escaping black hole increase with the central density until $10^5$ ${\rm M}_{\odot}{\rm pc}^{-3}$ because of the shorter relaxation time, higher encounter rates at the center \citep[see also figure 8 in][]{2015ApJ...800....9M}. The binary fraction of escaping black holes reaches 30 per cent at $10^5$ ${\rm M}_{\odot}{\rm pc}^{-3}$, which is consistent with the values from the literature \citep[e.g.,][]{2014MNRAS.440.2714B,2015ApJ...800....9M}. 

When the initial central density exceeds $10^6$ ${\rm M}_{\odot}{\rm pc}^{-3}$, on the other hand, the IMBH formation plays a more important role. In this density range, the in-cluster merger is the most major factor of vanishing the black hole population. As shown in Fig. \ref{F6}, the massive IMBH seeds formed through the runaway stellar collisions can consume a significant fraction of stellar mass black holes in the system for their mass buildup. The number of escaping black holes decreases with the central density, especially by the fast IMBH formation cases. Moreover, the fraction of binaries among the escaping black holes also decreases by a significant fraction. At the vicinity of the IMBH, black holes are rather captured by the IMBH instead of forming stellar mass black hole binaries, and the binary of the IMBH and a black hole can eject other single black holes via the strong gravitational interactions \citep{2014MNRAS.444...29L}. Therefore, the black hole population at the vicinity of the IMBH is quickly depleted. For stellar mass black hole binaries segregated from the outer region, exchange encounters with the IMBH can lead to the capture of one black hole and the ejection of its companion \citep{2007MNRAS.376L..29G}. However, once a binary black hole has safely escaped from the strong gravity of the IMBH, such binary should be a very hard one and will merge within a Hubble time as shown in Fig. \ref{F7}. 

\subsection{Modified correlation including IMBH cluster models}
\begin{figure}
  \centering
  \includegraphics[width=1.0\columnwidth]{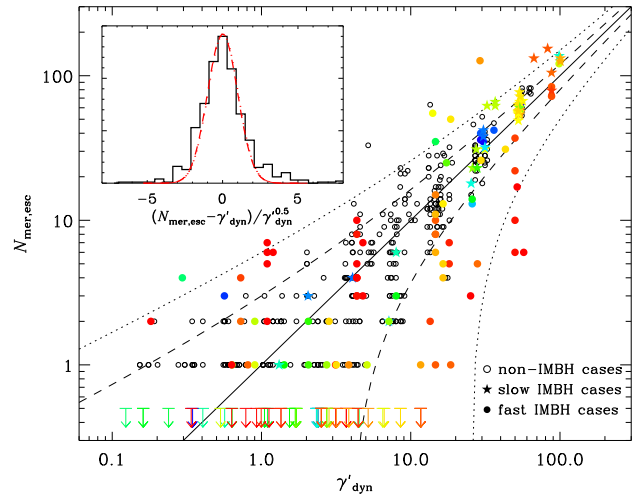}
  \caption{The modified correlation between the number of escaping merging binary black holes and the predicted numbers based on the equation \ref{E2}, for the entire set of simulation models. Colors and symbols are the same as Fig. \ref{F3} for IMBH cases, and non-IMBH cases are indicated as open circles. The dashed and dotted lines are 2 times and 5 times of the Poisson errors away from the diagonal line. The inset plot displays the histogram of differences of numbers to the predictions normalized by their Poisson errors. The red line shows a Gaussian distribution.}\label{F8}
\end{figure}
{In the previous sections, we described that the rapid formation and growth of the IMBH can reduce the binary black hole merger population significantly. From the Figs. \ref{F1} and \ref{F7}, we can infer that the impact of IMBHs via the fast formation becomes strong when the initial central density exceeds $10^6$ ${\rm M}_{\odot}{\rm pc}^{-3}$. Therefore, the correlation formula from the equation \ref{E1} can be modified by taking a factor into account as
\begin{equation}\label{E2}
\gamma'_{\rm dyn}\equiv\gamma_{\rm dyn}\times \Big(1+\frac{\rho_{\rm c,0}^2}{\tilde{\rho}_{\rm c}^2}\Big)^{-\frac{1}{2}},
\end{equation}
where $\rho_{\rm c,0}$ and $\tilde{\rho}_{\rm c}$ are the initial central density and a fitting parameter, respectively. We find $\tilde{\rho}_{\rm c}=10^{5.92}$ ${\rm M}_{\odot}{\rm pc}^{-3}$ for the best fitting. The equation gives $\gamma'_{\rm dyn} \cong \gamma_{\rm dyn}$ for $\rho_{\rm c,0} \ll \tilde{\rho}_{\rm c}$ and $\gamma'_{\rm dyn} \cong \gamma_{\rm dyn}(\tilde{\rho}_{\rm c}/\rho_{\rm c,0})$ for $\rho_{\rm c,0} \gg \tilde{\rho}_{\rm c}$.
Fig. \ref{F8} shows the correlation between the number of escaping merging binary black holes for the entire set of models and the expected numbers driven by the modified formula in equation \ref{E2}. The distribution of models is slightly broader than that of non-IMBH cases from Fig. \ref{F2}. In the inset of Fig. \ref{F8}, we present a histogram of the difference between the number of mergers and the prediction, which is then normalized by the square-root of the predicted numbers that corresponds to the Poisson error. It follows well a Gaussian distribution with some excesses in the tails, which is due to the stochastic effects of the formation and growth of IMBHs. As shown in Fig. \ref{F1}, non, slow and fast IMBH cluster models are not clearly separated in the range of initial central density between $10^4$ and $10^6$ ${\rm M}_{\odot}{\rm pc}^{-3}$. In addition, the degree of reduction for fast IMBH models can be determined by the final IMBH mass but highly unpredictable with an order of magnitude uncertainty (see Fig. \ref{F4}).

It is very challenging to predict the formation and growth of IMBHs and the consequent reduction of merger production for individual globular clusters directly from their initial conditions. However, the accumulation of binary black hole mergers from a huge number of globular clusters will follow the prediction from the equation \ref{E2}. \citet{2018MNRAS.480.5645H} has explored the merger rate density in the local Universe by using various assumptions concerning the properties of initial globular cluster systems such as distributions of the clusters' initial mass and half-mass radius}. In order to examine the effects of the IMBH formation on the local merger rate density, we repeat the same calculation with the modified correlation formula. We take $W_0=6$ as a representative value, and the initial globular cluster mass function following a \citet{1976ApJ...203..297S} function with a minimum mass $10^3$ \msun{} and a cut-off mass $10^{6.5}$ \msun. According to these assumptions\footnote{We only consider the `small' size distribution from \citet{2018MNRAS.480.5645H}, based on \citet{2012A&A...543A...8M}. Due to the much lower density for the `large' size distribution \citep{2004A&A...416..537L}, there is no discrepancy in the local merger rate density with the IMBH correction.}, the local merger rate density can be reduced by a factor of $\sim$2 by considering the effects of IMBH formation.

We note that the prescription and assumptions for the formation and growth of IMBHs used in this survey might be somewhat unrealistic \citep{2017MNRAS.464.3090A}. The IMBH mass growth efficiency is set to 100 per cent and the recoil kick after gravitational wave mergers is ignored. However, the mass growth efficiency for mergers between IMBH and stellar mass black holes can be 95 per cent according to the current gravitational wave detections. The mass growth of seed IMBHs via tidal captures also can be very efficient \citep{2017MNRAS.467.4180S}. Moreover, the merger recoil kick can be negligible for IMBH-stellar mass black hole mergers if the mass ratio $q<0.1$ \citep{2018MNRAS.481.2168M}. Although the assumptions used in the survey can affect the early growth and retention probability of seed IMBHs, the effect will be less significant once IMBHs have become sufficiently massive. We emphasize that the main purpose of this study is to investigate the effect of IMBHs on the formation and mergers of stellar mass black hole binaries, but not to provide the prediction of the overall detection rate changed by IMBHs in the cosmological context since our understanding on the formation and growth of IMBHs in globular clusters is still very limited. \\

\section{Summary and conclusion}
In this paper, we have explored a large set of Monte-Carlo cluster simulation survey results to investigate the effects of the formation and growth of intermediate mass black holes (IMBHs) in globular clusters on the population of stellar mass black hole binaries, in the perspective of their subsequent escapes and mergers by gravitational wave emission which is one of the main targets of ground based gravitational wave detectors. 

According to recent numerical studies, it has been confirmed that the production rate of binary black hole mergers from globular clusters is well correlated with the total mass and stellar density at the beginning. However, we have revealed that the production rate is significantly reduced when the host globular cluster forms an IMBH. The degree of reduction increases up to two orders of magnitude if the IMBH mass at present-day exceeds $10^4$ \msun{} and/or the final IMBH mass fraction compared to the initial total mass of host globular cluster becomes larger than 1 per cent. On the other hand, if the IMBH mass is not massive enough or its seed black hole forms later than 1 Gyr, then the production rate of black hole binary mergers is similar to that for non-IMBH cases. It is interesting that this discrepancy also corresponds to the different IMBH formation scenario, the fast and slow formation scenarios by \citet{2015MNRAS.454.3150G}. 
{ We found that the IMBH formation via the fast scenario frequently happens with a very large initial central density. The fraction of the final IMBH mass compared to the initial total cluster mass also depends on the initial central density.}

We checked that about 90 per cent of binary black holes merging within a Hubble time escape before IMBHs form and grow. This would make us able to draw a conclusion that the formation of an IMBH and the production of stellar mass black hole binaries compete each other in the beginning of clusters' dynamical evolution. Once an IMBH forms inside a globular cluster, stellar mass black holes rather merge into the IMBH or are ejected out as single black holes by strong gravitational interactions between stars at vicinity and the IMBH. Even existing black hole binaries segregated from the cluster outskirt are disrupted instead of becoming tightened. Thus, the presence of IMBHs leads to the significant reduction of binary black hole merger population via not only decreasing the number of escaping black holes but also by ejecting black holes as single objects. For the most extreme cases with the initial central density over $10^7$ ${\rm M}_{\odot}{\rm pc}^{-3}$, only one third of black holes can escape from the system and 10 per cent of escaping black holes are in binaries on average.

{ Based on the survey results, the correlation between the number of escaping merging binary black holes from \citet{2018MNRAS.480.5645H} has been corrected by applying a factor which is expressed by the initial central density. We then estimated the overall merger rate density in the local Universe and found that it can be reduced by a factor of 2 for given initial cluster mass function, cluster size distribution and the dimensionless concentration parameter.}
In order to apply our findings for the impact of IMBHs to the overall detection rate estimation, however, future confirmation of the presence of IMBHs and better knowledge of their populations in globular clusters by, such as, tidal disruption events and intermediate mass ratio inspirals will be needed \citep[see e.g.][]{2018ApJ...867..119F}. 

\section*{ACKNOWLEDGEMENTS}
We would thank the anonymous referee for the constructive suggestions that improved our manuscript. This research was supported by Basic Science Research Program through the National Research Foundation of Korea (NRF) funded by the Ministry of Education (No. 2020R1I1A1A01051827). 
AA acknowledges support from the Carl Tryggers Foundation for Scientific Research through the grant CTS 17:113 and from the Swedish Research Council through the grant 2017-04217. 
MG was partially supported by the Polish National Science Center (NCN) through the grant UMO-2016/23/B/ST9/02732.
AH was supported by NCN grant UMO-2016/20/S/ST9/00162.
SJY acknowledges support by the Mid-career Researcher Program (No.
2019R1A2C3006242) and the SRC Program (the Center for Galaxy Evolution
Research; No. 2017R1A5A1070354) through NRF.


\begin{thebibliography}{}
\bibitem[Aarseth(2012)]{2012MNRAS.422..841A} Aarseth, S.~J.\ 2012, \mnras, 422, 841 
\bibitem[Abbott et al.(2016)]{2016PhRvL.116f1102A} Abbott, B.~P., et al.\ 2016, Physical Review Letters, 116, 061102 
\bibitem[Abbott et al.(2017)]{2017PhRvL.119p1101A} Abbott, B.~P., et al.\ 2017, Physical Review Letters, 119, 161101 
\bibitem[\protect\citeauthoryear{Abbott et al.}{2019}]{2019PhRvX...9c1040A} Abbott B.~P., et al., 2019, PhRvX, 9, 031040
\bibitem[\protect\citeauthoryear{Abbott et al.}{2020}]{2020ApJ...892L...3A} Abbott B.~P., et al., 2020, ApJL, 892, L3
\bibitem[Amaro-Seoane \& Chen(2016)]{2016MNRAS.458.3075A} Amaro-Seoane P., Chen X.\ 2016, \mnras, 458, 3075 
\bibitem[\protect\citeauthoryear{Anderson \& van der Marel}{2010}]{2010ApJ...710.1032A} Anderson J., van der Marel R.~P., 2010, ApJ, 710, 1032
\bibitem[Antonini \& Perets(2012)]{2012ApJ...757...27A} Antonini, F., Perets, H.~B.\ 2012, \apj, 757, 27 
\bibitem[\protect\citeauthoryear{Antonini \& Gieles}{2020}]{2020MNRAS.492.2936A} Antonini F., Gieles M., 2020, MNRAS, 492, 2936
\bibitem[\protect\citeauthoryear{Antonini, Gieles \& Gualandris}{2019}]{2019MNRAS.486.5008A} Antonini F., Gieles M., Gualandris A., 2019, MNRAS, 486, 5008
\bibitem[\protect\citeauthoryear{Arca Sedda, Askar \& Giersz}{2018}]{2018MNRAS.479.4652A} Arca Sedda M., Askar A., Giersz M., 2018, MNRAS, 479, 4652
\bibitem[\protect\citeauthoryear{Arca Sedda et al.}{2019}]{2019arXiv190811375A} Arca Sedda M., et al., 2019, arXiv, arXiv:1908.11375
\bibitem[\protect\citeauthoryear{Arca Sedda et al.}{2020}]{2020ApJ...894..133A} Arca Sedda M., Mapelli M., Spera M., Benacquista M., Giacobbo N., 2020, ApJ, 894, 133
\bibitem[\protect\citeauthoryear{Askar, Arca Sedda \& Giersz}{2018}]{2018MNRAS.478.1844A} Askar A., Arca Sedda M., Giersz M., 2018, MNRAS, 478, 1844
\bibitem[Askar et al.(2017a)]{2017MNRAS.464L..36A} Askar A., Szkudlarek M., Gondek-Rosi{\'n}ska D., Giersz M.,   Bulik T.,\ 2017a, \mnras, 464, L36
\bibitem[\protect\citeauthoryear{Askar et al.}{2017b}]{2017MNRAS.464.3090A} Askar A., Bianchini P., de Vita R., Giersz M., Hypki A., Kamann S., 2017b, MNRAS, 464, 3090
\bibitem[Bae, Kim \& Lee(2014)]{2014MNRAS.440.2714B} Bae, Y.-B., Kim, C., Lee, H.~M.\ 2014, \mnras, 440, 2714 
\bibitem[Bae et al.(2017)]{2017PhRvD..96h4009B} Bae, Y.-B., Lee, H.~M., Kang, G., Hansen, J.\ 2017, \prd, 96, 084009 \bibitem[Banerjee(2017)]{2017MNRAS.467..524B} Banerjee S.,\ 2017, \mnras, 467, 524 
\bibitem[Banerjee, Baumgardt \& Kroupa(2010)]{2010MNRAS.402..371B} Banerjee, S., Baumgardt, H., Kroupa, P.\ 2010, \mnras, 402, 371 
\bibitem[Baumgardt, Makino \& Ebisuzaki(2004a)]{2004ApJ...613.1133B} Baumgardt H., Makino J.,   Ebisuzaki T.,\ 2004a, \apj, 613, 1133 
\bibitem[Baumgardt, Makino \& Ebisuzaki(2004b)]{2004ApJ...613.1143B} Baumgardt H., Makino J.,   Ebisuzaki T.,\ 2004b, \apj, 613, 1143 
\bibitem[\protect\citeauthoryear{Baumgardt, Makino \& Hut}{2005}]{2005ApJ...620..238B} Baumgardt H., Makino J., Hut P., 2005, ApJ, 620, 238
\bibitem[Belczynski et al.(2007)]{2007ApJ...662..504B} Belczynski, K., Taam, R.~E., Kalogera, V., Rasio, F.~A., Bulik, T.\ 2007, \apj, 662, 504 
\bibitem[\protect\citeauthoryear{Belczynski et al.}{2016}]{2016ApJ...819..108B} Belczynski K., et al., 2016, ApJ, 819, 108
\bibitem[\protect\citeauthoryear{Belczynski et al.}{2020}]{2020A&A...636A.104B} Belczynski K., et al., 2020, A\&A, 636, A104
\bibitem[Belczynski, Kalogera \& Bulik(2002)]{2002ApJ...572..407B} Belczynski K., Kalogera V.,   Bulik T.,\ 2002, \apj, 572, 407 
\bibitem[\protect\citeauthoryear{Bouffanais et al.}{2019}]{2019ApJ...886...25B} Bouffanais Y., Mapelli M., Gerosa D., Di Carlo U.~N., Giacobbo N., Berti E., Baibhav V., 2019, ApJ, 886, 25
\bibitem[\protect\citeauthoryear{Chatterjee et al.}{2017}]{2017ApJ...836L..26C} Chatterjee S., Rodriguez C.~L., Kalogera V., Rasio F.~A., 2017, ApJL, 836, L26
\bibitem[\protect\citeauthoryear{Choksi et al.}{2019}]{2019ApJ...873..100C} Choksi N., Volonteri M., Colpi M., Gnedin O.~Y., Li H., 2019, ApJ, 873, 100
\bibitem[\protect\citeauthoryear{Di Carlo et al.}{2019}]{2019MNRAS.487.2947D} Di Carlo U.~N., Giacobbo N., Mapelli M., Pasquato M., Spera M., Wang L., Haardt F., 2019, MNRAS, 487, 2947
\bibitem[\protect\citeauthoryear{Di Carlo et al.}{2020}]{2020MNRAS.tmp.2334D} Di Carlo U.~N., Mapelli M., Giacobbo N., Spera M., Bouffanais Y., Rastello S., Santoliquido F., et al., 2020, MNRAS.tmp, doi:10.1093/mnras/staa2286
\bibitem[\protect\citeauthoryear{de Vita et al.}{2017}]{2017MNRAS.467.4057D} de Vita R., Trenti M., Bianchini P., Askar A., Giersz M., van de Ven G., 2017, MNRAS, 467, 4057
\bibitem[Dominik et al.(2012)]{2012ApJ...759...52D} Dominik, M., Belczynski, K., Fryer, C., et al.\ 2012, \apj, 759, 52 
\bibitem[Downing et al.(2011)]{2011MNRAS.416..133D} Downing, J.~M.~B., Benacquista, M.~J., Giersz, M., Spurzem, R.\ 2011, \mnras, 416, 133 
\bibitem[\protect\citeauthoryear{Fragione et al.}{2018}]{2018ApJ...867..119F} Fragione G., Leigh N.~W.~C., Ginsburg I., Kocsis B., 2018, ApJ, 867, 119
\bibitem[Fregeau et al.(2004)]{2004MNRAS.352....1F} Fregeau J.~M., Cheung P., Portegies Zwart S.~F., Rasio F.~A.,\ 2004, \mnras, 352, 1 
\bibitem[Fujii, Tanikawa \& Makino(2017)]{2017PASJ...69...94F} Fujii M.~S., Tanikawa A., Makino J.,\ 2017, \pasj, 69, 94 
\bibitem[Fukushige \& Heggie(2000)]{2000MNRAS.318..753F} Fukushige T., Heggie D.~C.,\ 2000, \mnras, 318, 753 
\bibitem[\protect\citeauthoryear{Gebhardt, Rich \& Ho}{2005}]{2005ApJ...634.1093G} Gebhardt K., Rich R.~M., Ho L.~C., 2005, ApJ, 634, 1093
\bibitem[\protect\citeauthoryear{Gerssen et al.}{2002}]{2002AJ....124.3270G} Gerssen J., van der Marel R.~P., Gebhardt K., Guhathakurta P., Peterson R.~C., Pryor C., 2002, AJ, 124, 3270
\bibitem[Giacobbo, Mapelli \& Spera(2018)]{2018MNRAS.474.2959G} Giacobbo N., Mapelli M., Spera M.,\ 2018, \mnras, 474, 2959 
\bibitem[\protect\citeauthoryear{Giersz}{1998}]{1998MNRAS.298.1239G} Giersz M., 1998, MNRAS, 298, 1239
\bibitem[\protect\citeauthoryear{Giersz}{2001}]{2001MNRAS.324..218G} Giersz M., 2001, MNRAS, 324, 218
\bibitem[Giersz et al.(2013)]{2013MNRAS.431.2184G} Giersz M., Heggie D.~C., Hurley J.~R., Hypki A.,\ 2013, \mnras, 431, 2184 
\bibitem[Giersz et al.(2015)]{2015MNRAS.454.3150G} Giersz M., Leigh N., Hypki A., L{\"u}tzgendorf N., Askar A.,\ 2015, \mnras, 454, 3150 
\bibitem[\protect\citeauthoryear{Giersz et al.}{2019}]{2019MNRAS.487.2412G} Giersz M., Askar A., Wang L., Hypki A., Leveque A., Spurzem R., 2019, MNRAS, 487, 2412
\bibitem[\protect\citeauthoryear{Giersz \& Spurzem}{2003}]{2003MNRAS.343..781G} Giersz M., Spurzem R., 2003, MNRAS, 343, 781
\bibitem[Gond{\'a}n et al.(2018)]{2018ApJ...860....5G} Gond{\'a}n L., Kocsis B., Raffai P., Frei Z.\ 2018, \apj, 860, 5
\bibitem[\protect\citeauthoryear{Greene, Strader \& Ho}{2019}]{2019arXiv191109678G} Greene J.~E., Strader J., Ho L.~C., 2019, arXiv, arXiv:1911.09678
\bibitem[\protect\citeauthoryear{Gualandris \& Portegies Zwart}{2007}]{2007MNRAS.376L..29G} Gualandris A., Portegies Zwart S., 2007, MNRAS, 376, L29
\bibitem[\protect\citeauthoryear{H{\'e}non}{1971a}]{1971Ap&SS..14..151H} H{\'e}non M.~H., 1971, Ap\&SS, 14, 151
\bibitem[\protect\citeauthoryear{H{\'e}non}{1971b}]{1971Ap&SS..13..284H} H{\'e}non M., 1971, Ap\&SS, 13, 284
\bibitem[\protect\citeauthoryear{Hobbs et al.}{2005}]{2005MNRAS.360..974H} Hobbs G., Lorimer D.~R., Lyne A.~G., Kramer M., 2005, MNRAS, 360, 974
\bibitem[Hong \& Lee(2015)]{2015MNRAS.448..754H} Hong J., Lee H.~M.,\ 2015, \mnras, 448, 754 
\bibitem[\protect\citeauthoryear{Hong et al.}{2018}]{2018MNRAS.480.5645H} Hong J., Vesperini E., Askar A., Giersz M., Szkudlarek M., Bulik T., 2018, MNRAS, 480, 5645
\bibitem[Hurley, Pols \& Tout(2000)]{2000MNRAS.315..543H} Hurley J.~R., Pols O.~R., Tout C.~A.,\ 2000, \mnras, 315, 543 
\bibitem[Hurley, Tout \& Pols(2002)]{2002MNRAS.329..897H} Hurley J.~R., Tout C.~A., Pols O.~R.,\ 2002, \mnras, 329, 897 
\bibitem[Hypki \& Giersz(2013)]{2013MNRAS.429.1221H} Hypki A., Giersz M.,\ 2013, \mnras, 429, 1221 
\bibitem[\protect\citeauthoryear{Kimball, Berry \& Kalogera}{2020}]{2020RNAAS...4....2K} Kimball C., Berry C., Kalogera V., 2020, RNAAS, 4, 2
\bibitem[King(1966)]{1966AJ.....71...64K} King I.~R.,\ 1966, \aj, 71, 64 
\bibitem[\protect\citeauthoryear{Kremer et al.}{2020}]{2020arXiv200610771K} Kremer K., Spera M., Becker D., Chatterjee S., Di Carlo U.~N., Fragione G., Rodriguez C.~L., et al., 2020, arXiv, arXiv:2006.10771
\bibitem[\protect\citeauthoryear{Kroupa}{1995a}]{1995MNRAS.277.1491K} Kroupa P., 1995a, MNRAS, 277, 1491
\bibitem[Kroupa(1995b)]{1995MNRAS.277.1507K} Kroupa P.,\ 1995b, \mnras, 277,  1507
\bibitem[Kroupa(2001)]{2001MNRAS.322..231K} Kroupa P.,\ 2001, \mnras, 322, 231
\bibitem[\protect\citeauthoryear{Kumamoto, Fujii \& Tanikawa}{2019}]{2019MNRAS.486.3942K} Kumamoto J., Fujii M.~S., Tanikawa A., 2019, MNRAS, 486, 3942
\bibitem[Larsen(2004)]{2004A&A...416..537L} Larsen S.~S.,\ 2004, \aap, 416, 537 
\bibitem[\protect\citeauthoryear{Lanzoni et al.}{2013}]{2013ApJ...769..107L} Lanzoni B., et al., 2013, ApJ, 769, 107\bibitem[Leigh et al.(2014)]{2014MNRAS.444...29L} Leigh N.~W.~C., L{\"u}tzgendorf N., Geller A.~M., et al.,\ 2014, \mnras, 444, 29 
\bibitem[\protect\citeauthoryear{Lin et al.}{2018}]{2018NatAs...2..656L} Lin D., et al., 2018, NatAs, 2, 656
\bibitem[\protect\citeauthoryear{L{\"u}tzgendorf et al.}{2011}]{2011A&A...533A..36L} L{\"u}tzgendorf N., Kissler-Patig M., Noyola E., Jalali B., de Zeeuw P.~T., Gebhardt K., Baumgardt H., 2011, A\&A, 533, A36
\bibitem[\protect\citeauthoryear{Madrid et al.}{2017}]{2017MNRAS.470.1729M} Madrid J.~P., Leigh N.~W.~C., Hurley J.~R., Giersz M., 2017, MNRAS, 470, 1729
\bibitem[\protect\citeauthoryear{Mapelli et al.}{2019}]{2019MNRAS.487....2M} Mapelli M., Giacobbo N., Santoliquido F., Artale M.~C., 2019, MNRAS, 487, 2
\bibitem[Marks \& Kroupa(2012)]{2012A&A...543A...8M} Marks M., Kroupa P.,\ 2012, \aap, 543, A8 
\bibitem[\protect\citeauthoryear{Miller \& Hamilton}{2002}]{2002MNRAS.330..232C} Miller M.~C., Hamilton D.~P., 2002, MNRAS, 330, 232
\bibitem[\protect\citeauthoryear{Morawski et al.}{2018}]{2018MNRAS.481.2168M} Morawski J., Giersz M., Askar A., Belczynski K., 2018, MNRAS, 481, 2168
\bibitem[Morscher et al.(2015)]{2015ApJ...800....9M} Morscher, M., Pattabiraman, B., Rodriguez, C., Rasio, F.~A., Umbreit, S.\ 2015, \apj, 800, 9 
\bibitem[O'Leary et al.(2006)]{2006ApJ...637..937O} O'Leary, R.~M., Rasio, F.~A., Fregeau, J.~M., Ivanova, N., O'Shaughnessy, R.\ 2006, \apj, 637, 937 
\bibitem[O'Leary, Kocsis, \& Loeb(2009)]{2009MNRAS.395.2127O} O'Leary, R.~M., Kocsis, B., Loeb, A.\ 2009, \mnras, 395, 2127 
\bibitem[Park et al.(2017)]{2017MNRAS.469.4665P} Park D., Kim C., Lee H.~M., Bae Y.-B., Belczynski K.,\ 2017, \mnras, 469, 4665
\bibitem[Portegies Zwart \& McMillan(2000)]{2000ApJ...528L..17P} Portegies Zwart, S.~F., McMillan, S.~L.~W.\ 2000, \apjl, 528, L17 
\bibitem[Portegies Zwart \& McMillan(2002)]{2002ApJ...576..899P} Portegies Zwart S.~F., McMillan S.~L.~W.,\ 2002, \apj, 576, 899 
\bibitem[\protect\citeauthoryear{Portegies Zwart et al.}{2004}]{2004Natur.428..724P} Portegies Zwart S.~F., Baumgardt H., Hut P., Makino J., McMillan S.~L.~W., 2004, Natur, 428, 724
\bibitem[\protect\citeauthoryear{Rodriguez \& Antonini}{2018}]{2018ApJ...863....7R} Rodriguez C.~L., Antonini F., 2018, ApJ, 863, 7
\bibitem[Rodriguez, Chatterjee \& Rasio(2016)]{2016PhRvD..93h4029R} Rodriguez C.~L., Chatterjee S., Rasio F.~A.,\ 2016, \prd, 93, 084029
\bibitem[Rodriguez et al.(2015)]{2015PhRvL.115e1101R} Rodriguez, C.~L., Morscher, M., Pattabiraman, B., et al.\ 2015, Physical Review Letters, 115, 051101 
\bibitem[Rodriguez et al.(2016)]{2016ApJ...824L...8R} Rodriguez, C.~L., Haster, C.-J., Chatterjee, S., Kalogera, V., Rasio, F.~A.\ 2016, \apjl, 824, L8 
\bibitem[\protect\citeauthoryear{Rodriguez et al.}{2018}]{2018PhRvD..98l3005R} Rodriguez C.~L., et al., 2018, PhRvD, 98, 123005
\bibitem[\protect\citeauthoryear{Rodriguez et al.}{2020}]{2020ApJ...896L..10R} Rodriguez C.~L., et al., 2020, ApJL, 896, L10
\bibitem[Sakurai et al.(2017)]{2017MNRAS.472.1677S} Sakurai Y., Yoshida N., Fujii M.~S., Hirano S.,\ 2017, \mnras, 472, 1677 
\bibitem[\protect\citeauthoryear{Sakurai, Yoshida \& Fujii}{2019}]{2019MNRAS.484.4665S} Sakurai Y., Yoshida N., Fujii M.~S., 2019, MNRAS, 484, 4665
\bibitem[\protect\citeauthoryear{Samsing et al.}{2020}]{2020PhRvD.101l3010S} Samsing J., D'Orazio D.~J., Kremer K., Rodriguez C.~L., Askar A., 2020, PhRvD, 101, 123010
\bibitem[\protect\citeauthoryear{Samsing \& D'Orazio}{2018}]{2018MNRAS.481.5445S} Samsing J., D'Orazio D.~J., 2018, MNRAS, 481, 5445
\bibitem[\protect\citeauthoryear{Samsing, Hamers \& Tyles}{2019}]{2019PhRvD.100d3010S} Samsing J., Hamers A.~S., Tyles J.~G., 2019, PhRvD, 100, 043010
\bibitem[Schechter(1976)]{1976ApJ...203..297S} Schechter P.,\ 1976, \apj, 203, 297 
\newpage
\bibitem[\protect\citeauthoryear{Spera \& Mapelli}{2017}]{2017MNRAS.470.4739S} Spera M., Mapelli M., 2017, MNRAS, 470, 4739
\bibitem[\protect\citeauthoryear{Stodolkiewicz}{1982}]{1982AcA....32...63S} Stodolkiewicz J.~S., 1982, AcA, 32, 63
\bibitem[\protect\citeauthoryear{Stodolkiewicz}{1986}]{1986AcA....36...19S} Stodolkiewicz J.~S., 1986, AcA, 36, 19
\bibitem[\protect\citeauthoryear{Stone, K{\"u}pper \& Ostriker}{2017}]{2017MNRAS.467.4180S} Stone N.~C., K{\"u}pper A.~H.~W., Ostriker J.~P., 2017, MNRAS, 467, 4180
\bibitem[\protect\citeauthoryear{Stone \& Leigh}{2019}]{2019Natur.576..406S} Stone N.~C., Leigh N.~W.~C., 2019, Natur, 576, 406
\bibitem[\protect\citeauthoryear{Tremou et al.}{2018}]{2018ApJ...862...16T} Tremou E., et al., 2018, ApJ, 862, 16
\bibitem[Trenti et al.(2007)]{2007MNRAS.374..857T} Trenti M., Ardi E., Mineshige S., Hut P.,\ 2007, \mnras, 374, 857 
\bibitem[\protect\citeauthoryear{van der Marel \& Anderson}{2010}]{2010ApJ...710.1063V} van der Marel R.~P., Anderson J., 2010, ApJ, 710, 1063
\bibitem[Wang et al.(2016)]{2016MNRAS.458.1450W} Wang, L., Spurzem, R., Aarseth, S., et al.\ 2016, \mnras, 458, 1450 
\end{thebibliography}
\end{document}